\documentclass[useAMS,usenatbib,usegraphicx]{mn2e}

\renewcommand{\vec}[1]{\bmath{#1}}

\newcommand{\pd}[2]{\frac{\partial #1}{\partial #2}}
\newcommand{\DS}{\displaystyle}

\newcommand{\gtrsim}{{\ga}}
\newcommand{\lesssim}{{\la}}
\newcommand{\rhorest}{{\rho}} 
\newcommand{\hg}{{h}} 
\newcommand{\pg}{{p}} 

\title[Equation of state in RMHD]
      {Equation of State in Relativistic Magnetohydrodynamics:
       variable versus constant adiabatic index}

\author[A. Mignone and J.C. McKinney]{A. Mignone$^{1,2}$
 \thanks{E-mail:mignone@to.astro.it(AM);jmckinney@cfa.harvard.edu(JCM)} and Jonathan C. McKinney$^{3}$\footnotemark[1]\\
  $^{1}$INAF Osservatorio Astronomico di Torino, 10025 Pino Torinese, Italy\\
  $^{2}$Dipartimento di Fisica Generale dell'Universit\`a,
  Via Pietro Giuria 1, I-10125 Torino, Italy \\
  $^{3}$Institute for Theory and Computation, Center for
  Astrophysics, Harvard University, 60 Garden St.,
  Cambridge, MA, 02138}

\begin{document}
\date{Accepted 2007 April 12. Received 2007 April 12; in original form 2007 January 25}

\pagerange{\pageref{firstpage}--\pageref{lastpage}} \pubyear{2007}

\maketitle

\label{firstpage}

\begin{abstract}
  The role of the equation of state for a perfectly conducting,
  relativistic magnetized fluid is the main subject of this work.  The
  ideal constant $\Gamma$-law equation of state, commonly adopted in a
  wide range of astrophysical applications, is compared with a more
  realistic equation of state that better approximates the
  single-specie relativistic gas. The paper focus on three different
  topics.  First, the influence of a more realistic equation of state
  on the propagation of fast magneto-sonic shocks is investigated.
  This calls into question the validity of the constant $\Gamma$-law
  equation of state in problems where the temperature of the gas
  substantially changes across hydromagnetic waves.  Second, we
  present a new inversion scheme to recover primitive variables (such
  as rest-mass density and pressure) from conservative ones that
  allows for a general equation of state and avoids catastrophic
  numerical cancellations in the non-relativistic and
  ultrarelativistic limits.  Finally, selected numerical tests of
  astrophysical relevance (including magnetized accretion flows around
  Kerr black holes) are compared using different equations of state.
  Our main conclusion is that the choice of a realistic equation of
  state can considerably bear upon the solution when transitions from
  cold to hot gas (or viceversa) are present. Under these circumstances, a
  polytropic equation of state can significantly endanger the
  solution.
\end{abstract}


\section{Introduction}
%
%
%
%
%

Recent developments in numerical hydrodynamics have made a breach in
the understanding of astrophysical phenomena commonly associated with
relativistic magnetized plasmas.  Existence of such flows has nowadays
been largely witnessed by observations indicating superluminal motion
in radio loud active galactic nuclei and galactic binary systems, as
well as highly energetic events occurring in proximity of X-ray
binaries and super-massive black holes.  Strong evidence suggests that
the two scenarios may be closely related and that the production of
relativistic collimated jets results from magneto-centrifugal
mechanisms taking place in the inner regions of rapidly spinning
accretion disks \citep{MKU01}.

Due to the high degree of nonlinearity present in the equations of
relativistic magnetohydrodynamics (RMHD henceforth), analytical models are
often of limited applicability, relying on simplified assumptions
of time independence and/or spatial symmetries.
For this reason, they are frequently superseded by numerical models
that appeal to a consolidated theory based on finite difference
methods and Godunov-type schemes.
The propagation of relativistic supersonic jets without magnetic
field has been studied, for instance, in the pioneering work of
\cite{vP93, DH94} and, subsequently, by
\cite{MMFIM97,HRHD98, AIMGM99, MYT04} and references therein.
Similar investigations in presence of poloidal and toroidal magnetic
fields have been carried on by \cite{NSCSM97, Koide97, K99} and more
recently by \cite{LAAM05, MMB05}.

The majority of analytical and numerical models, including the
aforementioned studies, makes extensive use of the polytropic equation of state
(EoS henceforth), for which the specific heat ratio is constant and equal
to $5/3$ (for a cold gas) or to $4/3$ (for a hot gas).
However, the theory of relativistic perfect gases \citep{Synge57}
teaches that, in the limit of negligible free path, the ratio of specific heats
cannot be held constant if consistency with the kinetic theory is to be required.
This was shown in an even earlier work by \cite{Taub48}, where a fundamental
inequality relating specific enthalpy and temperature was proved to
hold.

Although these results have been known for many decades, only few
investigators seem to have faced this important aspect.
\cite{DHO96} suggested, in the context of extragalactic jets,
the importance of self-consistently computing a variable adiabatic index
rather than using a constant one.
This may be advisable, for example, when the dynamics is regulated by
multiple interactions of shock waves, leading to the formation of
shock-heated regions in an initially cold gas.
Lately, \cite{SAMGM02} addressed similar issues by investigating the long term
evolution of jets with an arbitrary mixture of electrons, protons and
electron-positron pairs.
Similarly, \cite{MSTV04} considered thermally accelerated outflows in
proximity of compact objects by adopting a variable effective polytropic index
to account for transitions from non-relativistic to relativistic temperatures.
Similar considerations pertain to models of Gamma Ray Burst (GRB) engines
including accretion discs, which have an EoS that must account for a
combination of protons, neutrons, electrons,
positrons, and neutrinos, etc. and must include the effects of electron
degeneracy, neutronization, photodisintegration, optical depth of neutrinos,
etc. \citep{PWF99, DPN02, KM02, KNP05}. However, for the disk that is mostly
photodisintegrated and optically thin to neutrinos, a decent approximation
of such EoS is a variable $\Gamma$-law with $\Gamma=5/3$ when the temperature
is below $m_e c^2/k_b$ and $\Gamma=4/3$ when above $m_e c^2/k_b$ due to the
production of positrons at high temperatures that gives a relativistic plasma
(Broderick, McKinney, Kohri in prep.).
Thus, the variable EoS considered here may be a reasonable approximation of GRB
disks once photodisintegration has generated mostly free nuclei.

The additional complexity introduced by more elaborate EoS
comes at the price of extra computational cost since the
EoS is frequently used in the process of obtaining numerical
solutions, see for example, \cite{FK96}.
Indeed, for the Synge gas, the correct EoS does not have a simple
analytical expression and the thermodynamics of the fluid becomes
entirely formulated in terms of the modified Bessel functions.

Recently \citet[][ MPB henceforth]{MPB05} introduced, in the context of
relativistic non-magnetized flows, an approximate EoS
that differs only by a few percent from the theoretical one. The
advantage of this approximate EoS, earlier adopted by \cite{Mat71},
is its simple analytical representation.
A slightly better approximation, based on an analytical
expression, was presented by \cite{RCC06}.

In the present work we wish to discuss the role of the EoS in RMHD,
with a particular emphasis to the one proposed by MPB, properly
generalized to the context of relativistic magnetized flows.
Of course, it is still a matter of debate the extent to which
equilibrium thermodynamic principles can be correctly prescribed when
significant deviations from the single-fluid ideal approximation may hold
(e.g., non-thermal particle distributions, gas composition, cosmic ray
acceleration and losses, anisotropy, and so forth).
Nevertheless, as the next step in a logical course of action,
we will restrict our attention to a single aspect - namely the use of a
constant polytropic versus a variable one - and we will ignore the influence of
such non-ideal effects (albeit potentially important) on the EoS.


In \S\ref{sec:eqns}, we present the relevant equations and discuss the
properties of the new EoS versus the more restrictive constant
$\Gamma$-law EoS.  In \S\ref{sec:shock}, we consider the propagation
of fast magneto-sonic shock waves and solve the jump conditions across
the front using different EoS.  As we shall see, this
calls into question the validity of the constant $\Gamma$-law EoS in
problems where the temperature of the gas substantially changes across
hydromagnetic waves.
In \S\ref{sec:numerical}, we present numerical simulations
of astrophysical relevance such
as blast waves, axisymmetric jets, and magnetized accretion disks around
Kerr black holes.
A short survey of some existing models is conducted using
different EoS's in order to determine if significant interesting
deviations arise.
These results should be treated as a guide to some
possible avenues of research rather than as the definitive result
on any individual topic.
Results are summarized in \S\ref{sec:conclusion}.
In the Appendix, we present a description of the primitive variable inversion scheme.

\section{Relativistic MHD Equations}
\label{sec:eqns}
%
%
%
%
%

In this section we present the equations of motion for relativistic
MHD, discuss the validity of the ideal gas EoS as
applied to a perfect gas, and review an alternative EoS
that properly models perfect gases in both the hot (relativistic) and
cold (non-relativistic) regimes.

\subsection{Equations of Motion}
%
%
%

Our starting point are the relativistic MHD equations in conservative
form:
\begin{equation}\label{eq:CL}
  \pd{}{t}\left(\begin{array}{c}
    D       \\ \noalign{\medskip}
    \vec{m} \\ \noalign{\medskip}
    \vec{B} \\ \noalign{\medskip}
    E      \\
  \end{array}\right)
  +
  \nabla\cdot\left(\begin{array}{c}
    D\vec{v}  \\ \noalign{\medskip}
    w_t \gamma^2 \vec{v}\vec{v} - \vec{b}\vec{b} + Ip_t \\ \noalign{\medskip}
    \vec{v}\vec{B} - \vec{B}\vec{v} \\ \noalign{\medskip}
    \vec{m}
  \end{array}\right) = 0 \,,
\end{equation}
together with the divergence-free constraint $\nabla\cdot\vec{B} = 0$,
where $\vec{v}$ is the velocity, $\gamma$ is the Lorentz factor,
$w_t\equiv (\rho h + \pg + b^2)$ is the relativistic total
(gas+magnetic) enthalpy, $p_t=\pg+b^2/2$ is the total (gas+magnetic)
fluid pressure, $\vec{B}$ is the lab-frame field, and the field in the
fluid frame is given by
\begin{equation}
  b^\alpha = \gamma \{\vec{v}\cdot\vec{B},
                      \frac{B^i}{\gamma^2} + v^i(\vec{v}\cdot\vec{B}) \} ,
\end{equation}
with an energy density of
\begin{equation}
  |\vec{b}|^2 = \frac{|\vec{B}|^2}{\gamma^2} + (\vec{v}\cdot\vec{B})^2 .
\end{equation}
Units are chosen such that the speed of light is equal to one.
Notice that the fluxes entering in the induction equation
are the components of the electric field that, in the infinite
conductivity approximation, become
\begin{equation}
  \vec{\Omega} = -\vec{v}\times\vec{B} \;.
\end{equation}
The non-magnetic case is recovered by letting $\vec{B}\to 0$ in
the previous expressions.

The conservative variables are, respectively, the laboratory density
$D$, the three components of momentum $m_k$ and magnetic field $B_k$
and the total energy density $E$:
\begin{eqnarray}
  D   & = & \rhorest\gamma   \;,  \label{eq:cons_var_D}\\ \noalign{\medskip}
  m_k & = & (D \hg \gamma + |\vec{B}|^2)v_k - (\vec{v}\cdot\vec{B})B_k  \; ,
  \label{eq:cons_var_m}\\ \noalign{\medskip}
  E   & = & \DS D \hg\gamma  - \pg
  + \frac{|\vec{B}|^2}{2} + \frac{|\vec{v}|^2 |\vec{B}|^2 - (\vec{v}\cdot\vec{B})^2}{2}
  \label{eq:cons_var_E}\;,
\end{eqnarray}

The specific enthalpy $h$ and internal energy $\epsilon$ of the gas
are related by
\begin{equation}
  h = 1 + \epsilon + \frac{p}{\rho}\,,
\end{equation}
and an additional equation of state relating two thermodynamical
variables (e.g. $\rho$ and $\epsilon$) must be specified for proper closure.
This is the subject of the next section.

Equations (\ref{eq:cons_var_D})--(\ref{eq:cons_var_E}) are
routinely used in numerical codes to recover conservative
variables from primitive ones (e.g., $\rho$, $\vec{v}$, $p$ and
$\vec{B}$). The inverse relations cannot be cast in closed
form and require the solution of one or more nonlinear
equations. \cite{Noble06} review several methods of inversion
for the constant $\Gamma$-law, for which $\rho \epsilon = p/(\Gamma - 1)$.
We present, in Appendix \ref{sec:inversion}, the details of a new inversion
procedure suitable for a more general EoS.

\subsection{Equation of State}
%
%
%

\begin{figure}
  \begin{center}
    \includegraphics[width=.5\textwidth]{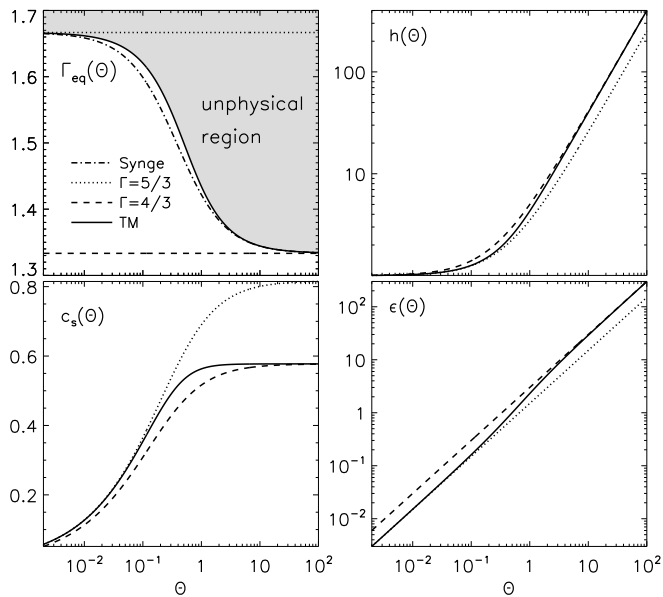}
    \caption{Equivalent $\Gamma$ (top left), specific enthalpy (top
      right), sound speed (bottom left) and specific internal energy
      (bottom right) as functions of temperature $\Theta =
      p/\rhorest$.  Different lines correspond to the various
      EoS mentioned the text: the ideal
      $\Gamma=5/3$-law (dotted line), ideal $\Gamma=4/3$-law (dashed
      line), TM EoS (solid line).  For clarity the Synge-gas
      (dashed-dotted line) has been plotted only in the top left
      panel, where the ``unphysical region" marks the area where
      Taub's inequality is not fulfilled.}
    \label{fig:eosplot}
  \end{center}
\end{figure}

Proper closure to the conservation law (\ref{eq:CL}) is required in
order to solve the equations.  This is achieved by specifying an
EoS relating thermodynamic quantities.  The theory of
relativistic perfect gases shows that the specific enthalpy is a
function of the temperature $\Theta = \pg/\rhorest$ alone and it takes
the form \citep{Synge57}
\begin{equation}\label{eq:RP}
  \hg = \frac{K_3(1/\Theta)}{K_2(1/\Theta)}\,,
\end{equation}
where $K_2$ and $K_3$ are, respectively, the order 2 and 3 modified
Bessel functions of the second kind.
Equation (\ref{eq:RP}) holds for a gas composed of material
particles with the same mass and in the limit of small free
path when compared to the sound wavelength.

Direct use of Eq. (\ref{eq:RP}) in numerical codes, however,
results in time-consuming algorithms and alternative
approaches are usually sought.
The most widely used and popular one relies on the choice of the
constant $\Gamma$-law EoS
\begin{equation}\label{eq:ID}
  \hg = 1 + \frac{\Gamma}{\Gamma - 1}\Theta\,,
\end{equation}
where $\Gamma$ is the constant specific heat ratio.
However, \cite{Taub48} showed that
consistency with the relativistic kinetic theory requires the
specific enthalpy $\hg$ to satisfy
\begin{equation}\label{eq:taub_ineq}
  \left(\hg - \Theta\right)
  \left(\hg - 4\Theta\right) \ge 1 \,,
\end{equation}
known as Taub's fundamental inequality.
Clearly the constant $\Gamma$-law EoS does not fulfill
(\ref{eq:taub_ineq}) for an arbitrary choice of $\Gamma$,
while (\ref{eq:RP}) certainly does.
This is better understood in terms of an equivalent $\Gamma_{\rm eq}$,
conveniently defined as
\begin{equation}
  \Gamma_{\rm eq} = \frac{h - 1}{h - 1 - \Theta} \,,
\end{equation}
and plotted in the top left panel of Fig. \ref{fig:eosplot} for
different EoS.
In the limit of low and high temperatures, the physically
admissible region is delimited, respectively, by
$\Gamma_{\rm eq} \le 5/3$ (for $\Theta\to0$)
and $\Gamma_{\rm eq} \le 4/3$ (for $\Theta\to\infty$).
Indeed, Taub's inequality is always fulfilled when
$\Gamma \le 4/3$ while it cannot be satisfied for
$\Gamma \ge 5/3$ for any positive value of the temperature.

In a recent paper, \cite{MPB05} showed that
if the equal sign is taken in Eq. (\ref{eq:taub_ineq}), an equation
with the correct limiting values may be derived.
The resulting EoS ($TM$ henceforth), previously introduced by \cite{Mat71},
can be solved for the enthalpy, yielding
\begin{equation}\label{eq:TM}
  \hg = \frac{5}{2}\Theta + \sqrt{\frac{9}{4}\Theta^2 + 1}\,,
\end{equation}
or, using $\rho h= \rho + \rho\epsilon + p$ in (\ref{eq:taub_ineq}) with
the equal sign,
\begin{equation}\label{eq:TM_int_en}
  p = \frac{\rho\epsilon\left(\rho\epsilon + 2\rho\right)}{3\left(\rho\epsilon + \rho\right)}
    = \frac{\epsilon + 2}{\epsilon + 1}\,\frac{\rho\epsilon}{3}\,.
\end{equation}
Direct evaluation of $\Gamma_{\rm eq}$ using (\ref{eq:TM}) shows
that the $TM$ EoS differs by less than $4\%$ from the theoretical value
given by the relativistic perfect gas EoS (\ref{eq:RP}).
The proposed EoS behaves closely to the $\Gamma=4/3$ law
in the limit of high temperatures, whereas reduces to the $\Gamma=5/3$
law in the cold gas limit. For intermediate temperatures, thermodynamical
quantities (such as specific internal energy, enthalpy and sound
speed) smoothly vary between the two limiting cases, as illustrated in
Fig. \ref{fig:eosplot}.
In this respect, Eq. (\ref{eq:TM}) greatly improves over the constant
$\Gamma$-law EoS and, at the same time, offers ease of implementation over
Eq. (\ref{eq:RP}).
Since thermodynamics is frequently invoked during the numerical
solution of (\ref{eq:CL}), it is expected that direct implementation
of Eq. (\ref{eq:TM}) in numerical codes will result in faster and more
efficient algorithms.

Thermodynamical quantities such as sound speed and entropy are computed
from the $2^{\rm nd}$ law of thermodynamics,
\begin{equation}\label{eq:2nd_law}
  dS = \frac{dh}{\Theta} - d\log p \,,
\end{equation}
where $S$ is the entropy.
From the definition of the sound speed,
\begin{equation}
  c_s^2 \equiv \left.\pd{p}{e}\right|_S \,,
\end{equation}
and using $de = hd\rho$ (at constant $S$), one finds
the useful expression
\begin{equation}\label{eq:cspeed}
  c_s^2 = \frac{\Theta}{\hg}\frac{\dot{\hg}}{\dot{\hg} - 1} =
  \left\{\begin{array}{cc}
  \DS  \frac{\Gamma\Theta}{\hg}                         & \quad \textrm{$\Gamma$-law EoS} \,,\\ \noalign{\medskip}
  \DS  \frac{\Theta}{3\hg}\frac{5\hg-8\Theta} {\hg-\Theta}  & \quad \textrm{TM EoS} \,.
  \end{array}\right.
\end{equation}
where we set $\dot{\hg}= d\hg/d\Theta$.
In a similar way, direct integration of (\ref{eq:2nd_law}) yields
$S = k\log\sigma$ with
\begin{equation}\label{eq:entropy}
 \sigma =
  \left\{\begin{array}{cc}
  \DS  \frac{p}{\rho^\Gamma}               & \quad \textrm{$\Gamma$-law EoS} \,,\\ \noalign{\medskip}
  \DS  \frac{p}{\rho^{5/3}}(h-\Theta)  & \quad \textrm{TM EoS} \,.
\end{array}\right.
\end{equation}
with $h$ given by (\ref{eq:TM}).

\section{Propagation of Fast Magneto-sonic Shocks}
\label{sec:shock}
%
%
%

Motivated by the previous results, we now investigate the
role of the EoS on the propagation of magneto-sonic
shock waves.
To this end, we proceed by constructing a one-parameter
family of shock waves with different velocities, traveling in
the positive $x$ direction.
States ahead and behind the front are labeled with
$\vec{U}_0$ and $\vec{U}_1$, respectively, and are related
by the jump conditions
\begin{equation}\label{eq:jc}
  v_s\left[\vec{U}\right] = \left[\vec{F}(\vec{U})\right] \,,
\end{equation}
where $v_s$ is the shock speed and
$\left[q\right] = q_1 - q_0$ is the jump across the wave for
any quantity $q$.
The set of jump conditions (\ref{eq:jc}) may be reduced
\citep{L76} to the following five positive-definite
scalar invariants
\begin{equation}\label{eq:jump1}
  \left[J\right] = 0       \,,
\end{equation}
\begin{equation}\label{eq:jump2}
  \left[h\eta\right] = 0   \,,
\end{equation}
\begin{equation}\label{eq:jump3}
  \left[{\cal H}\right] = \left[\frac{\eta^2}{J^2} - \frac{b^2}{\rhorest^2}\right] =  0 \,,
\end{equation}
\begin{equation}\label{eq:jump4}
  J^2 + \frac{\left[p+b^2/2\right]}{\left[h/\rhorest\right]} = 0   \,,
\end{equation}
\begin{equation}\label{eq:jump5}
  \left[h^2\right]
  + J^2\left[\frac{h^2}{\rhorest^2}\right]
  + 2{\cal H}\left[p\right]
  + 2\left[b^2\frac{h}{\rhorest}\right] =0  \,,
\end{equation}
where
\begin{equation}\label{eq:massflux}
  J = \rhorest\gamma\gamma_s(v_s - v^x) \,,
\end{equation}
is the mass flux across the shock, and
\begin{equation}
  \eta = -\frac{J}{\rhorest}(\vec{v}\cdot\vec{B}) + \frac{\gamma_s}{\gamma}B_x \,.
\end{equation}
Here $\gamma_s$ denotes the Lorentz factor of the shock.
Fast or slow magneto-sonic shocks may be discriminated
through the condition $\alpha_0>\alpha_1>0$ (for the formers)
or $\alpha_1<\alpha_0<0$ (for the latters), where
$\alpha = h/\rhorest - {\cal H}$.

We consider a pre-shock state characterized by a cold
($p_0=10^{-4}$) gas with density $\rhorest = 1$.
Without loss of generality, we choose a frame
of reference where the pre-shock velocity normal to
the front vanishes, i.e., $v_{x0} = 0$.
Notice that, for a given shock speed, $J^2$ can be computed from
the pre-shock state and thus one has to solve only
Eqns. (\ref{eq:jump2})--(\ref{eq:jump5}).

\subsection{Purely Hydrodynamical Shocks}
%

\begin{figure}
  \begin{center}
    \includegraphics[width=.5\textwidth]{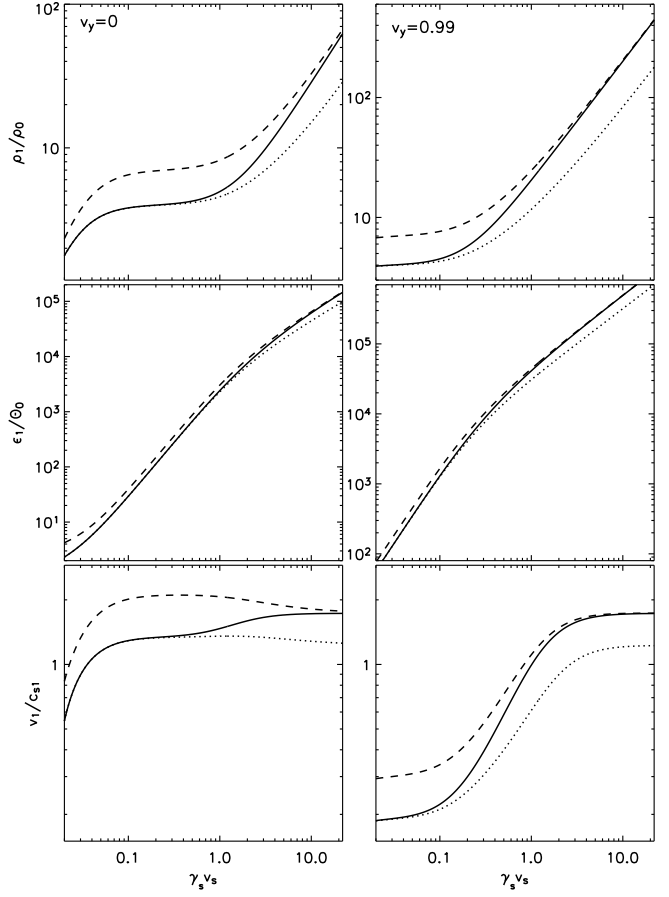}
    \caption{Compression ratio (top panels), internal energy (middle
      panels) and downstream Mach number (bottom panels) as functions
      of the shock four-velocity $\gamma_sv_s$.  The profiles give the
      solution to the shock equation for the non magnetic case.  Plots
      on the left have zero tangential velocity ahead of the front,
      whereas plots on right are initialized with $v_{y0} = 0.99$.
      Axis spacing is logarithmic. Solid, dashed and dotted lines
      correspond to the solutions obtained with the TM EoS and the
      $\Gamma=4/3$ and $\Gamma=5/3$ laws, respectively.}
    \label{fig:rhdshock}
  \end{center}
\end{figure}
In the limit of vanishing magnetic field, only
Eqns. (\ref{eq:jump4}) and (\ref{eq:jump5}) need
to be solved. Since $J^2$ is given, the problem
simplifies to the $2\times 2$ nonlinear system of equations
\begin{eqnarray}
  J^2 + \frac{\left[p\right]}{\left[\hg/\rhorest\right]} & = & 0 \,,\\
  \left[p\right]\left(\frac{h_1}{\rhorest_1} + \frac{\hg_0}{\rhorest_0}\right) - \left[\hg^2\right] &=& 0\,.
\end{eqnarray}

We solve the previous equations starting from $v_s=0.2$,
for which we were able to provide a sufficiently close guess
to the downstream state.
Once the $p_1$ and $\rhorest_1$ have been found, we repeat the
process by slowly increasing the shock velocity $v_s$ and using the previously
converged solution as the initial guess for the new value of $v_s$.

Fig. \ref{fig:rhdshock} shows the compression ratio, post-shock
internal energy $\epsilon_1$ and Mach number $v_1/c_{s1}$ as
functions of the shock four velocity $v_s\gamma_s$.
For weakly relativistic shock speeds and vanishing tangential
velocities (left panels), density and pressure jumps
approach the classical (i.e. non relativistic) strong
shock limit at $\gamma_sv_s\approx 0.1$, with the density ratio being
$4$ or $7$ depending on the value of $\Gamma$ ($5/3$ or $4/3$, respectively).
The post-shock temperature keeps non-relativistic
values ($\Theta\ll 1$) and the TM EoS behaves closely
to the $\Gamma = 5/3$ case, as expected.

With increasing shock velocity, the compression ratio does not saturate
to a limiting value (as in the classical case) but keeps growing at
approximately the same rate for the constant $\Gamma$-law EoS cases, and
more rapidly for the TM EoS.
This can be better understood by solving the jump conditions
in a frame of reference moving with the shocked material and then
transforming back to our original system. Since thermodynamics quantities
are invariant one finds that, in the limit $h_1\gg h_0 \approx 1$, the internal
energy becomes $\epsilon_1 = \gamma_1 - 1$ and the compression ratio takes
the asymptotic value
\begin{equation}
  \frac{\rhorest_1}{\rhorest_0} = \gamma_1 + \frac{\gamma_1 + 1}{\Gamma - 1} \,,
\end{equation}
when the ideal EoS is adopted.
Since $\gamma_1$ can take arbitrarily large values, the downstream density
keeps growing indefinitely.
At the same time, internal energy behind the shock rises faster than the rest-mass
energy, eventually leading to a thermodynamically relativistic configuration.
In absence of tangential velocities (left panels in Fig. \ref{fig:rhdshock}),
this transition starts at moderately high shock velocities
($\gamma_sv_s \ga 1$) and culminates when the shocked gas heats up to relativistic
temperatures ($\Theta\sim 1\div 10$) for $\gamma_sv_s \ga10$.
In this regime the TM EoS departs from the $\Gamma=5/3$ case and merges
on the $\Gamma = 4/3$ curve.
For very large shock speeds, the Mach number tends to the asymptotic
value $(\Gamma-1)^{-1/2}$, regardless of the frame of reference.


Inclusion of tangential velocities (right panels in Fig.
\ref{fig:rhdshock}) leads to an increased mass flux
($J^2\propto\gamma^2_0$) and, consequently, to higher post-shock
pressure and density values.
Still, since pressure grows faster than density, temperature in the post-shock flow strains to relativistic values even for slower shock
velocities and the TM EoS tends to the $\Gamma=4/3$ case at even smaller shock
velocities ($\gamma_sv_s \ga 2$).

Generally speaking, at a given shock velocity, density and pressure in the shocked gas
attain higher values for lower $\Gamma_\rmn{eq}$. Downstream temperature, on the
other hand, follows the opposite trend being higher as
$\Gamma_\rmn{eq}\to 5/3$ and lower when $\Gamma_\rmn{eq}\to4/3$.

\subsection{Magnetized Shocks}
%

\begin{figure}
  \begin{center}
    \includegraphics[width=.5\textwidth]{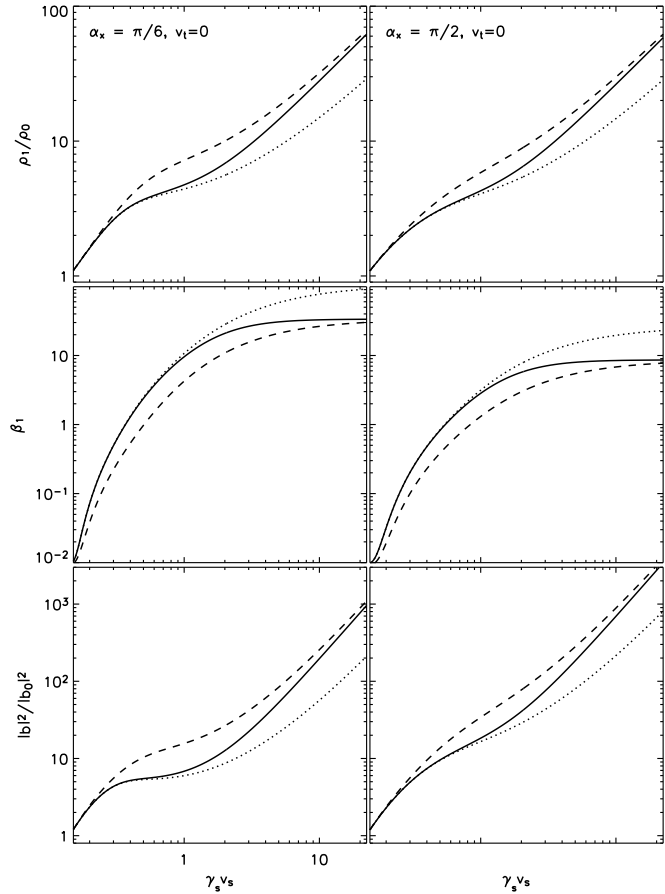}
    \caption{Compression ratio (top), downstream plasma $\beta$
      (middle) and magnetic field strength (bottom)
      as function of the shock four-velocity $\gamma_sv_s$
      with vanishing tangential component
      of the velocity. The magnetic field makes an angle
      $\pi/6$ (left) and $\pi/2$ (right) with the shock
      normal. The meaning of the different lines is the same
      as in Fig. \ref{fig:rhdshock}.}
    \label{fig:rmhdshock_noyvel}
  \end{center}
\end{figure}

\begin{figure}
  \begin{center}
    \includegraphics[width=.5\textwidth]{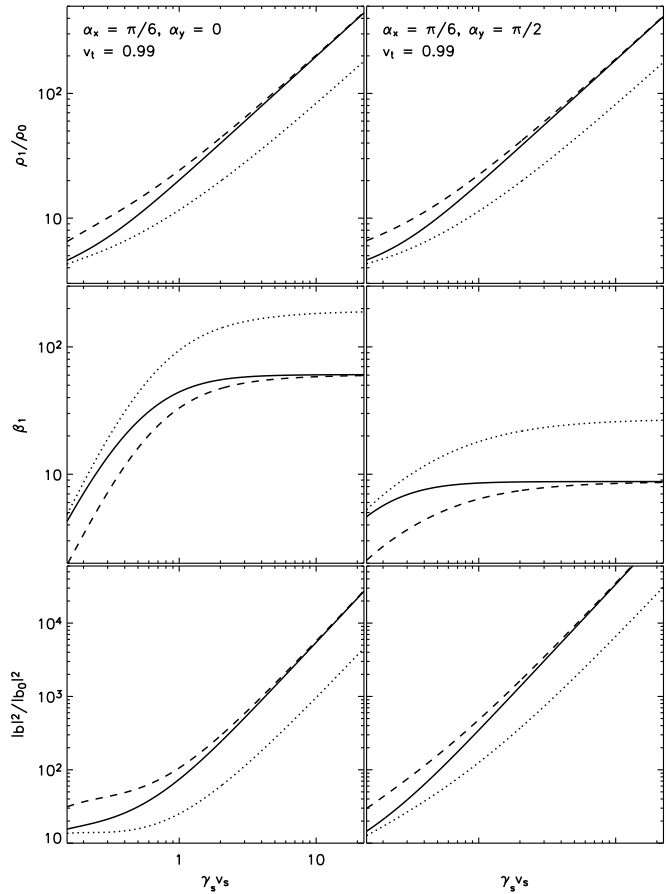}
    \caption{Density ratio (top), downstream plasma $\beta$
      (middle) and magnetic field strength (bottom)
      as function of $\gamma_sv_s$ when the tangential component
      of the upstream velocity is $v_t = 0.99$.
      The magnetic field and the shock normal form an
      angle $\pi/6$. The tangential components of magnetic
      field and velocity are aligned (left) and orthogonal
      (right).Different lines have the same meaning as in Fig.
      \ref{fig:rhdshock}.}
    \label{fig:rmhdshock_yvel_30}
  \end{center}
\end{figure}

\begin{figure}
  \begin{center}
    \includegraphics[width=.5\textwidth]{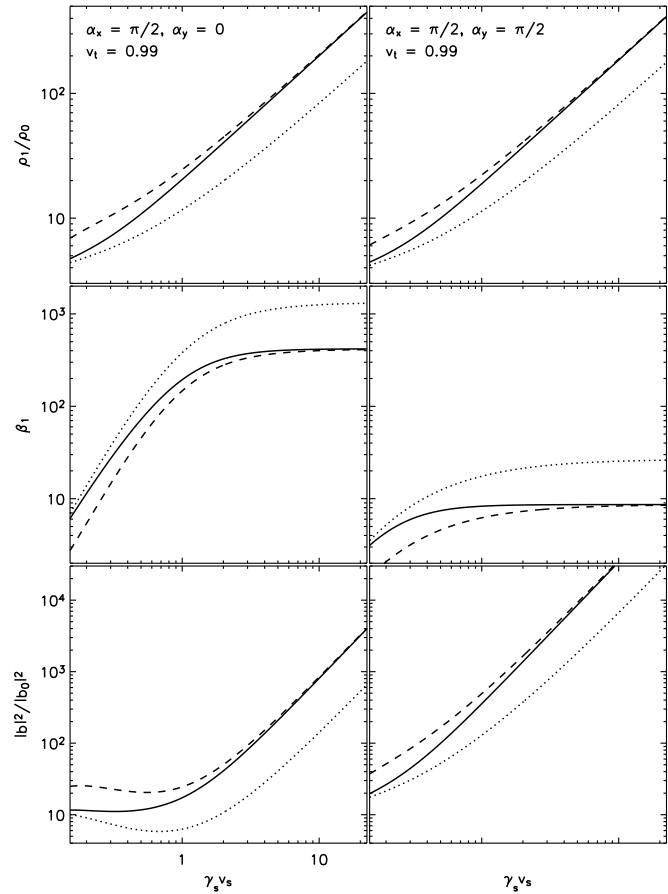}
    \caption{Density contrast (top), plasma $\beta$ (middle) and
      magnetic field strength (bottom) for $v_t = 0.99$.
      The magnetic field is purely transverse and
      aligned with the tangential component of velocity
      on the left, while it is orthogonal on the right.
      Different lines have the same meaning as in Fig.
      \ref{fig:rhdshock}.}
    \label{fig:rmhdshock_yvel_90}
  \end{center}
\end{figure}

In presence of magnetic fields, we solve the $3\times 3$ nonlinear system
given by Eqns. (\ref{eq:jump3}), (\ref{eq:jump4}) and (\ref{eq:jump5}),
and directly replace $\eta_1 = \eta_0 h_0/h_1$ with the aid of
Eq. (\ref{eq:jump2}).
The magnetic field introduces three additional parameters,
namely, the thermal to magnetic pressure ratio ($\beta \equiv 2p/b^2$)
and the orientation of the magnetic field with respect to the shock front
and to the tangential velocity.
This is expressed by the angles $\alpha_x$ and $\alpha_y$ such that
$B_x = |B|\cos\alpha_x$, $B_y = |B|\sin\alpha_x\cos\alpha_y$,
$B_z = |B|\sin\alpha_x\sin\alpha_y$.
We restrict our attention to the case of a strongly magnetized
pre-shock flow with $\beta_0\equiv 2p_0/b_0^2 = 10^{-2}$.

Fig. \ref{fig:rmhdshock_noyvel} shows the density, plasma $\beta$ and
magnetic pressure ratios versus shock velocity for $\alpha_x = \pi/6$
(left panels) and $\alpha_x = \pi/2$ (perpendicular shock, right panels).
Since there is no tangential velocity, the solution depends on one angle
only ($\alpha_x$) and the choice of $\alpha_y$ is irrelevant.
For small shock velocities ($\gamma_sv_s \la 0.4$), the front is
magnetically driven with density and pressure jumps attaining lower
values than the non-magnetized counterpart.
A similar behavior is found in classical MHD \citep{JT64}.
Density and magnetic compression ratios across the shock
reach the classical values around $\gamma_sv_s \approx 1$
(rather than $\gamma_sv_s\approx 0.1$ as in the non-magnetic
case) and increase afterwards.
The magnetic pressure ratio grows faster for the perpendicular
shock, whereas internal energy and density show little dependence
on the orientation angle $\alpha_x$.
As expected, the TM EoS mimics the constant $\Gamma=5/3$ case
at small shock velocities.
At $\gamma_sv_s \la 0.46$, the plasma $\beta$ exceeds unity and
the shock starts to be pressure-dominated.
In other words, thermal pressure eventually overwhelms the Lorentz
force and the shock becomes pressure-driven for velocities of the
order of $v_s\approx 0.42$.
When $\gamma_sv_s\ga 1$, the internal energy begins to become comparable to
the rest mass energy ($c^2$) and the behavior of the TM EoS detaches
from the $\Gamma=5/3$ curve and slowly joins the $\Gamma=4/3$ case.
The full transition happens in the limit of strongly relativistic
shock speeds, $\gamma_sv_s \la 10$.

Inclusion of transverse velocities in the right state
affects the solution in a way similar to the non-magnetic case.
Relativistic effects play a role already at small
velocities because of the increased inertia of the pre-shock
state introduced by the upstream Lorentz factor.
For $\alpha_x=\pi/6$ (Fig. \ref{fig:rmhdshock_yvel_30}), the compression
ratio does not drop to small values and keeps growing becoming
even larger ($\la 400$) than the previous case when $v_t = 0$.
The same behavior is reflected on the growth of magnetic pressure
that, in addition, shows more dependence on the relative
orientation of the velocity and magnetic field projections in
the plane of the front. When $\alpha_y = \pi/2$, indeed,
magnetic pressure attains very large values
($b^2/b_0^2 \la 10^4$, bottom right panel in
Fig. \ref{fig:rmhdshock_yvel_30}). Consequently, this is
reflected in a decreased post-shock plasma $\beta$.
For the TM EoS, the post-shock properties of the flow
begin to resemble the $\Gamma=4/3$ behavior at lower shock
velocities than before, $\gamma_sv_s\approx 2\div3$.
Similar considerations may be done for the case of a perpendicular
shock ($\alpha_x = \pi/2$, see Fig. \ref{fig:rmhdshock_yvel_90}),
although the plasma $\beta$ saturates to larger values thus
indicating larger post-shock pressures.
Again, the maximum increase in magnetic pressure occurs when
the velocity and magnetic field are perpendicular.

\section{Numerical Simulations}
\label{sec:numerical}
%
%
%
%
%

With the exception of very simple flow configurations, the solution of
the RMHD fluid equations must be carried out numerically.  This allows
an investigation of highly nonlinear regimes and complex interactions
between multiple waves. We present some examples of astrophysical
relevance, such as the propagation of one dimensional blast waves, the
propagation of axisymmetric jets, and the evolution of magnetized
accretion disks around Kerr black holes. Our goal is to outline the
qualitative effects of varying the EoS for some interesting astrophysical
problems rather than giving detailed results on any individual topic.

Direct numerical integration of Eq. (\ref{eq:CL}) has been
achieved using the PLUTO code \citep{PLUTO} in \S\ref{sec:blast}, \S\ref{sec:jet}
and HARM \citep{GmKT03} in \S\ref{sec:kerr}.
The new primitive variable inversion scheme presented in Appendix \ref{sec:inversion}
has been implemented in both codes and the results presented in
\S\ref{sec:blast} were used for code validation.
The novel inversion scheme offers the advantage of being suitable for a
more general EoS and avoiding catastrophic cancellation in the
non-relativistic and ultrarelativistic limits.

\subsection{Relativistic Blast Waves}
\label{sec:blast}
%
%
%

\begin{figure}
  \begin{center}
    \includegraphics[width=.5\textwidth]{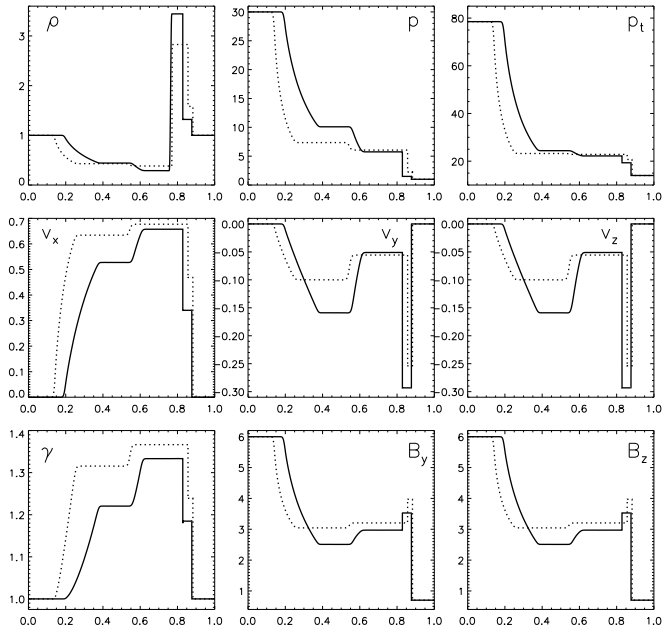}
    \caption{Solution to the mildly relativistic
      blast wave (problem 1) at t = 0.4.
      From left to right, the different profiles give
      density, thermal pressure, total pressure (top panels),
      the three components of velocity (middle
      panel) and magnetic fields (bottom panels).
      Computations with the TM EoS and constant
      $\Gamma=5/3$ EoS are shown using solid and dotted
      lines, respectively.}
    \label{fig:blast2}
  \end{center}
\end{figure}

\begin{figure}
  \begin{center}
    \includegraphics[width=.5\textwidth]{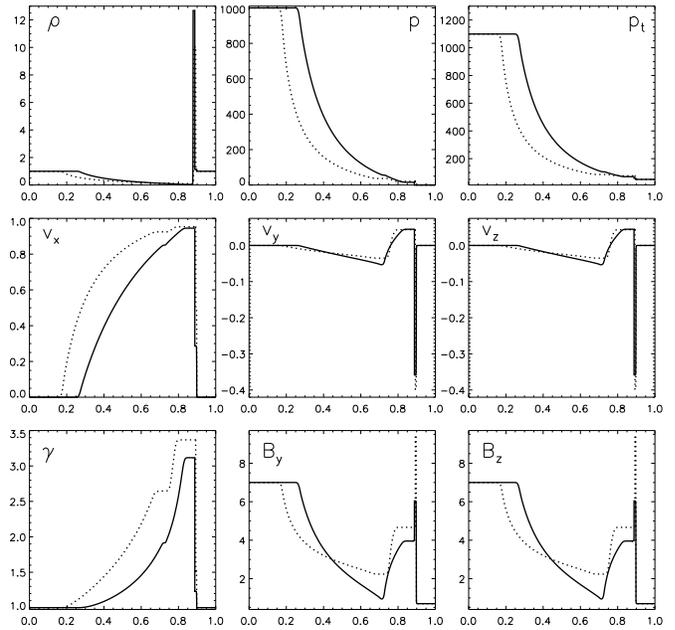}
    \caption{Solution to the strong relativistic
      blast wave (problem 2) at t = 0.4.
      From left to right, the different profiles give
      density, thermal pressure, total pressure (top panels),
      the three components of velocity (middle
      panel) and magnetic fields (bottom panels).
      Computations with the TM EoS and constant
      $\Gamma=5/3$ EoS are shown using solid and dotted
      lines, respectively.}
    \label{fig:blast3}
  \end{center}
\end{figure}

A shock tube consists of a sharp discontinuity separating two constant
states. In what follows we will be considering the one dimensional
interval $[0,1]$ with a discontinuity placed at $x=0.5$.  For the
first test problem, states to the left and to the right of the
discontinuity are given by $(\rhorest, p, B_y, B_z)_L = (1,30,6,6)$
for the left state and $(\rhorest,p,B_y, B_z)_R = (1,1,0.7,0.7)$ for
the right state.  This results in a mildly relativistic configuration
yielding a maximum Lorentz factor of $1.3\le\gamma\le1.4$.  The second
test consists of a left state given by $(\rhorest, p, B_y, B_z)_L =
(1,10^3,7,7)$ and a right state $(\rhorest,p,B_y, B_z)_R =
(1,0.1,0.7,0.7)$. This configuration involves the propagation of a
stronger blast wave yielding a more relativistic configuration
($3\le\gamma\le 3.5$).  For both states, we use a base grid with $800$
zones and $6$ levels of refinement (equiv. resolution = $800\cdot
2^6$) and evolve the solution up to $t=0.4$.

Computations carried with the ideal EoS with $\Gamma = 5/3$ and the TM
EoS are shown in Fig. \ref{fig:blast2} and Fig. \ref{fig:blast3} for
the first and second shock tube, respectively.  From left to right,
the wave pattern is comprised of a fast and slow rarefactions, a
contact discontinuity and a slow and a fast shocks.  No rotational
discontinuity is observed.  Compared to the $\Gamma=5/3$ case, one can
see that the results obtained with the TM EoS show considerable
differences.  Indeed, waves propagate at rather smaller velocities and
this is evident at the head and the tail points of the left-going
magneto-sonic rarefaction waves.  From a simple analogy with the
hydrodynamic counterpart, in fact, we know that these points propagate
increasingly faster with higher sound speed.  Since the sound speed
ratio of the TM and $\Gamma=5/3$ is always less than one (see, for
instance, the bottom left panel in Fig. \ref{fig:eosplot}), one may
reasonably predict slower propagation speed for the Riemann fans when
the TM EoS is used.  Furthermore, this is confirmed by computations
carried with $\Gamma=4/3$ that shows even slower velocities.  Similar
conclusions can be drawn for the shock velocities.  The reason is that
the opening of the Riemann fan of the TM equation state is smaller
than the $\Gamma=5/3$ case, because the latter always over-estimates
the sound speed.  The higher density peak behind the slow shock
follows from the previous considerations and the conservation of mass
across the front.

\subsection{Propagation of Relativistic Jets}
\label{sec:jet}
%
%
%

\begin{figure}
  \begin{center}
    \includegraphics[width=.5\textwidth]{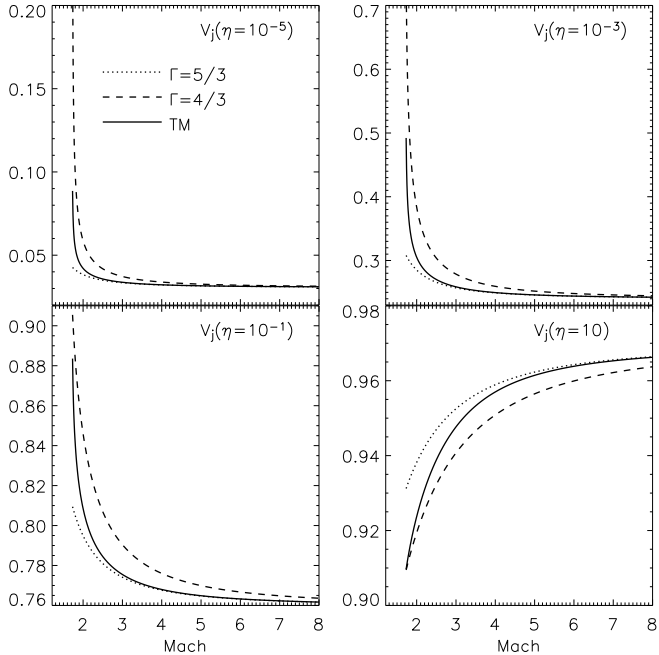}
    \caption{Jet velocity as a function of the Mach number for different
      values of the initial density contrast $\eta$.
      The beam Lorentz factor is the same for all plots,
      $\gamma_b=10$. Solid, dashed and dotted lines
      correspond to the solutions obtained with the TM EoS
      and the $\Gamma=4/3$ and $\Gamma=5/3$ laws, respectively.}
    \label{fig:jetvel}
  \end{center}
\end{figure}

\begin{figure}
  \begin{center}
    \includegraphics[width=.5\textwidth]{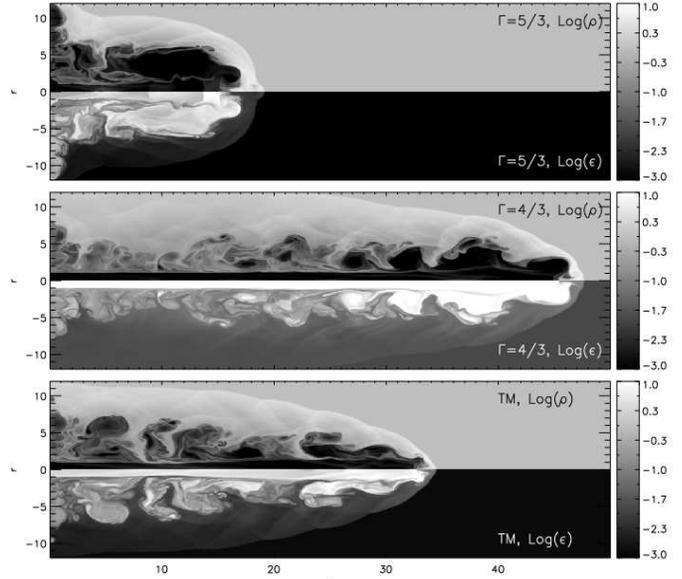}
    \caption{Computed results for the non magnetized jet at $t=90$ for the
      ideal EoS ($\Gamma=5/3$ and $\Gamma=4/3$, top and middle panels)
      and the TM EoS (bottom panel), respectively. The lower and upper
      half of each panels shows the gray-scale map of density and
      internal energy in logarithmic scale.}
    \label{fig:rhdjet}
  \end{center}
\end{figure}

\begin{figure}
  \begin{center}
    \includegraphics[width=.5\textwidth]{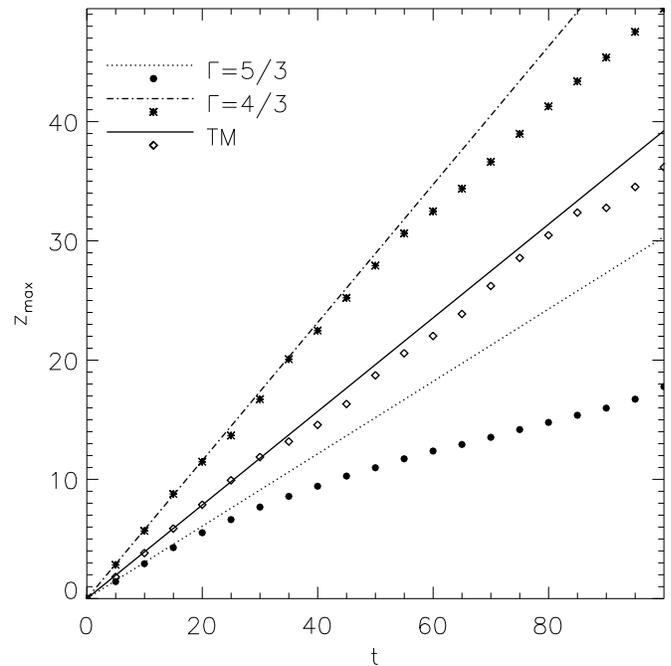}
    \caption{Position of the working surface as a function of time
      for $\Gamma=5/3$ (circles), $\Gamma=4/3$ (stars)
      and the TM EoS (diamonds). Solid, dotted and
      dashed lines gives the one-dimensional expectation.}
    \label{fig:rhdpos}
  \end{center}
\end{figure}

\begin{figure}
  \begin{center}
    \includegraphics[width=.5\textwidth]{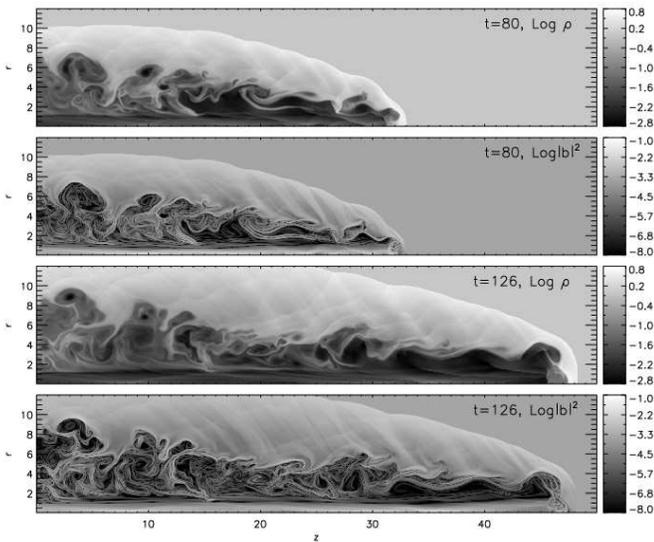}
    \caption{Density and magnetic field for the magnetized jet
      at $t=80$ (first and second panels from top) and at $t=126$
      (third and fourth panels). Computations were carried with
      $40$ zones per beam radius with the TM EoS.}
    \label{fig:rmhdjet}
  \end{center}
\end{figure}

Relativistic, pressure-matched jets are usually set up
by injecting a supersonic cylindrical beam with radius
$r_b$ into a uniform static ambient medium \citep[see, for instance,][]{MMFIM97}.
The dynamical and morphological properties of the jet and its
interaction with the surrounding are most commonly investigated
by adopting a three parameter set:
the beam Lorentz factor $\gamma_b$, Mach number $M_b = v_b/c_s$
and the beam to ambient density ratio $\eta=\rho_b/\rho_m$.
The presence of a constant poloidal magnetic field introduces a
fourth parameter $\beta_b = 2p_b/b^2$, which specifies
the thermal to magnetic pressure ratio.

\subsubsection{One Dimensional Models}
%

The propagation of the jet itself takes place at the velocity $V_j$,
defined as the speed of the working surface that separates
shocked ambient fluid from the beam material.
A one-dimensional estimate of $V_j$ (for vanishing magnetic fields) can be
derived from momentum flux balance in the frame of the working
surface \citep{MMFIM97}. This yields
\begin{equation}\label{eq:jetvel}
  V_j = \frac{\gamma_b\sqrt{\eta h_b/h_m}}{1 + \gamma_b\sqrt{\eta h_b/h_m}} \,,
\end{equation}
where $h_b$ and $h_m$ are the specific enthalpies of the beam and
the ambient medium, respectively. For given $\gamma_b$ and density
contrast $\eta$, Eq. (\ref{eq:jetvel}) may be regarded as a function
of the Mach number alone that uniquely specifies the pressure $p_b$
through the definitions of the sound speed, Eq. (\ref{eq:cspeed}).
For the constant $\Gamma$-law EoS the inversion is straightforward, whereas for
the TM EoS one finds, using the substitution $\Theta = 2/3\sinh x$,
\begin{equation}
  p_b = \eta\frac{2}{3}\sqrt{\frac{t^2_m}{1 - t_m^2}} \,,
\end{equation}
where $t_m$ satisfies the negative branch of the quadratic
equation
\begin{equation}
  t^2\left(15 - 6\frac{M_b^2}{v_b^2}\right)  +
  t  \left(24 - 10\frac{M_b^2}{v_b^2}\right) + 9 = 0 \,,
\end{equation}
with $t=\tanh x$.
In Fig. \ref{fig:jetvel} we show the jet velocity for increasing Mach
numbers (or equivalently, decreasing sound speeds) and different
density ratios $\eta = 10^{-5}, 10^{-3}, 10^{-1},1 0$. The Lorentz
beam factor is $\gamma_b = 10$.
Prominent discrepancies between the selected EoS arise at low Mach numbers,
where the relative variations of the jet speed between the constant
$\Gamma$ and the TM EoS's can be more than $50 \%$.
This regime corresponds to the case of a hot jet ($\Theta\approx 10$ in the
$\eta = 10^{-3}$ case) propagating into a cold ($\Theta \approx 10^{-3}$)
medium, for which neither the $\Gamma=4/3$ nor the $\Gamma=5/3$
approximation can properly characterize both fluids.

\subsubsection{Two Dimensional Models}
%

Of course, Eq. (\ref{eq:jetvel}) is strictly valid for one-dimensional
flows and the question remains as to whether similar conclusions
can be drawn in more than one dimension.
To this end we investigate, through numerical simulations, the propagation
of relativistic jets in cylindrical axisymmetric coordinates $(r,z)$.
We consider two models corresponding to different sets of
parameters and adopt the same computational domain $[0,12]\times[0,50]$
(in units of jet radius) with the beam being injected at the inlet region
($r \le 1$, $z=0$). Jets are in pressure equilibrium with the environment.

In the first model, the density ratio, beam Lorentz factor and Mach
number are given, respectively, by $\eta = 10^{-3}$, $\gamma_b = 10$
and $M_b = 1.77$. Magnetic fields are absent.  Integration are carried
at the resolution of $20$ zones per beam radius using the relativistic
Godunov scheme described in MPB. Computed results showing density and
internal energy maps at $t=90$ are given in Fig. \ref{fig:rhdjet} for
$\Gamma=5/3$, $\Gamma=4/3$ and the TM EoS.  The three different cases
differ in several morphological aspects, the most prominent one being
the position of the leading bow shock, $z\approx 18$ when
$\Gamma=5/3$, $z\approx 48$ for $\Gamma=4/3$ and $z\approx 33$ for the
TM EoS.  Smaller values of $\Gamma$ lead to larger beam internal
energies and therefore to an increased momentum flux, in agreement
with the one dimensional estimate (\ref{eq:jetvel}). This favors
higher propagation velocities and it is better quantified in Fig.
\ref{fig:rhdpos} where the position of the working surface is plotted
as a function of time and compared with the one dimensional estimate.
For the cold jet ($\Gamma=5/3$), the Mach shock exhibits a larger
cross section and is located farther behind the bow shock when
compared to the other two models. As a result, the jet velocity
further decreases promoting the formation of a thicker cocoon.  On the
contrary, the hot jet ($\Gamma=4/3$) propagates at the highest
velocity and the cocoon has a more elongated shape. The beam
propagates almost undisturbed and cross-shocks are weak.  Close to is
termination point, the beam widens and the jet slows down with hot
shocked gas being pushed into the surrounding cocoon at a higher rate.
Integration with the TM EoS reveals morphological and dynamical
properties more similar to the $\Gamma=4/3$ case, although the jet is
$\approx 40\%$ slower.  At $t=90$ the beam does not seem to decelerate
and its speed remains closer to the one-dimensional expectation.  The
cocoon develops a thinner structure with a more elongated conical
shape and cross shocks form in the beam closer to the Mach disk.

In the second case, we compare models C2-pol-1 and B1-pol-1 of
\cite{LAAM05} (corresponding to an ideal gas with $\Gamma = 5/3$ and
$\Gamma = 4/3$, respectively) with the TM EoS adopting the same
numerical scheme.  For this model, $\eta = 10^{-2}$, $v_b = 0.99$,
$M_b = 6$ and the ambient medium is threaded by a constant vertical
magnetic field, $B_z = \sqrt{2p_b}$.  Fig. \ref{fig:rmhdjet} shows the
results at $t=80$ and $t=126$, corresponding to the final integration
times shown in \cite{LAAM05} for the selected values of $\Gamma$.  For
the sake of conciseness, integration pertaining to the TM EoS only are
shown and the reader is reminded to the original work by \cite{LAAM05}
for a comprehensive description.  Compared to ideal EoS cases, the jet
shown here possesses morphological and dynamical properties
intermediate between the hot ($\Gamma=4/3$) and the cold
($\Gamma=5/3$) cases.  As expected, the jet propagates slower than in
model B1-pol-1 (hot jet), but faster than the cold one (C2-pol-1).
The head of the jet tends to form a hammer-like structure (although
less prominent than the cold case) towards the end of the integration,
i.e., for $t\ga 100$, but the cone remains more confined at previous
times.  Consistently with model C2-pol-1, the beam develops a series
of weak cross shocks and outgoing waves triggered by the interaction
of the flow with bent magnetic field lines.  Although the magnetic
field inhibits the formation of eddies, turbulent behavior is still
observed in cocoon, where interior cavities with low magnetic fields
are formed.  In this respect, the jet seems to share more features
with the cold case.

\subsection{Magnetized Accretion near Kerr Black Holes}
\label{sec:kerr}
%
%
%

\begin{figure}
\includegraphics[width=3.0in,clip]{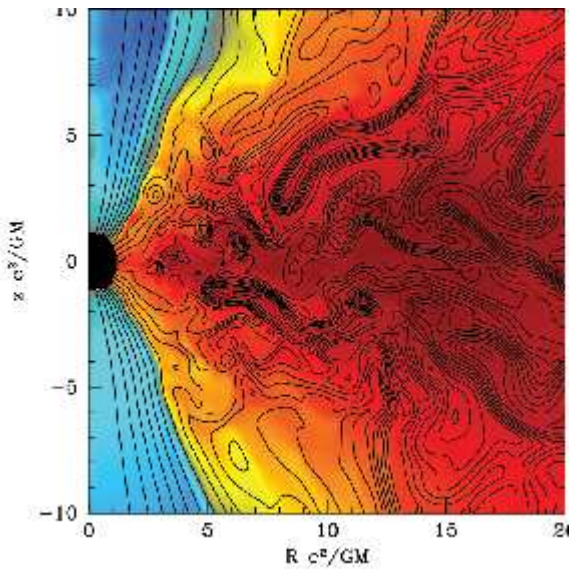}
  \begin{center}
    \caption{Magnetized accretion flow around a Kerr black hole for
      the ideal $\Gamma$-law EoS with $\Gamma=4/3$.  Shows the
      logarithm of the rest-mass density in colour from high (red) to
      low (blue) values.  The magnetic field has been overlayed.  This
      model demonstrates more vigorous turbulence and a thicker corona
      that leads to a more confined magnetized jet near the poles.}
    \label{fig:grmhddiskideal43}
  \end{center}
\end{figure}

\begin{figure}
\includegraphics[width=3.0in,clip]{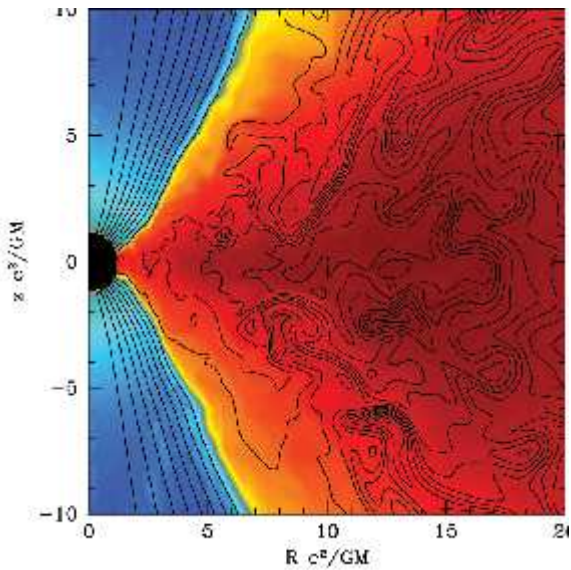}
  \begin{center}
    \caption{As in figure~\ref{fig:grmhddiskideal43} but for
    $\Gamma=5/3$.  Compared to the $\Gamma=4/3$ model, there is less
    vigorous turbulence and the corona is more sharply defined.}
    \label{fig:grmhddiskideal53}
  \end{center}
\end{figure}

\begin{figure}
\includegraphics[width=3.0in,clip]{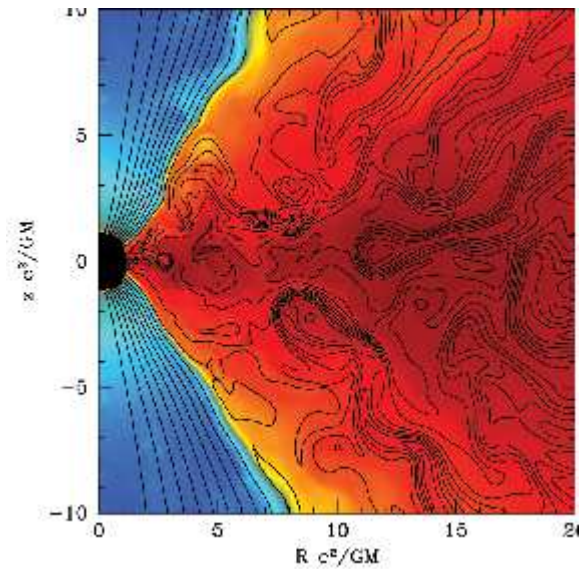}
  \begin{center}
    \caption{As in figure~\ref{fig:grmhddiskideal43} but for the TM
    EoS.  This EoS leads to turbulence that is less vigorous than in
    the $\Gamma=4/3$ model but more vigorous than in the $\Gamma=5/3$
    model.  Qualitatively the TM EoS leads to an accretion disk that
    behaves somewhere between the behavior of the $\Gamma=4/3$ and
    $\Gamma=5/3$ models.}
    \label{fig:grmhddisktm}
  \end{center}
\end{figure}

In this section we study time-dependent GRMHD numerical models of
black hole accretion in order to determine the effect of the EoS
on the behavior of the accretion disk, corona, and jet. We
study three models similar to the models studied by \citet{McKG04} for
a Kerr black hole with $a/M\approx 0.94$ and a disk with a scale
height ($H$) to radius ($R$) ratio of $H/R\sim 0.3$.  The constant
$\Gamma$-law EoS with $\Gamma=\{4/3,5/3\}$ and the TM EoS are used.
The initial torus solution is in hydrostatic equilibrium for the
$\Gamma$-law EoS, but we use the $\Gamma=5/3$ EoS as an initial
condition for the TM EoS.  Using the $\Gamma=4/3$ EoS as an initial
condition for the TM EoS did not affect the final quasi-stationary
behavior of the flow.  The simplest question to ask is which value of
$\Gamma$ will result in a solution most similar to the TM EoS model's
solution.

More advanced questions involve how the structure of the accretion
flow depends on the EoS. The previous results of this paper indicate
that the corona above the disk seen in the simulations
\citep{dev03,McKG04} will be most sensitive to the EoS since this
region can involve both non-relativistic and relativistic
temperatures.  The corona is directly involved is the production of a
turbulent, magnetized, thermal disk wind \citep{mn06a,mck06ff}, so the
disk wind is also expected to depend on the EoS.  The disk inflow near
the black hole has a magnetic pressure comparable to the gas pressure
\citep{McKG04}, so the EoS may play a role here and affect the flux of
mass, energy, and angular momentum into the black hole. The magnetized
jet associated with the~\citeauthor{bz77} solution seen in simulations
\citep{McKG04,mckinney2006} is not expected to depend directly on the
EoS, but may depend indirectly through the confining action of the
corona.  Finally, the type of field geometries observed in simulations
that thread the disk and corona \citep{hirose04,McK05} might depend on
the EoS through the effect of the stiffness (larger $\Gamma$ leads to
harder EoSs) of the EoS on the turbulent diffusion of magnetic fields.

Figs. \ref{fig:grmhddiskideal43}, \ref{fig:grmhddiskideal53} and
\ref{fig:grmhddisktm} show a snapshot of the accretion disk, corona,
and jet at $t\sim 1000GM/c^3$.  Overall the results are quite
comparable, as could be predicted since the $\Gamma=\{4/3,5/3\}$
models studied in \citet{McKG04} were quite similar.  For all models,
the field geometries allowed are similar to that found in
\citet{McK05}.  The accretion rate of mass, specific energy, and
specific angular momentum are similar for all models, so the EoS
appears to have only a small effect on the flow through the disk near
the black hole.

The most pronounced effect is that the soft EoS
($\Gamma=4/3$) model develops more vigorous turbulence due to the
non-linear behavior of the magneto-rotational instability (MRI) than
either the $\Gamma=5/3$ or TM EoSs.  This causes the coronae in the
$\Gamma=4/3$ model to be slightly thicker and to slightly more
strongly confine the magnetized jet resulting in a slight decrease in
the opening angle of the magnetized jet at large radii.  Also, the
$\Gamma=4/3$ model develops a fast magnetized jet at slightly smaller
radii than the other models.  An important consequence is that the
jet opening angle at large radii might depend sensitively on the
EoS of the material in the accretion disc corona.
This should be studied in future work.

\section{Conclusions}
\label{sec:conclusion}
%
%
%
%
%

The role of the EoS in relativistic magnetohydrodynamics
has been investigated both analytically and numerically. The equation
of state previously introduced by \cite{MPB05} (for non magnetized
flows) has been extended to the case where magnetic fields are present.
The proposed equation of state closely approximates the single-specie
perfect relativistic gas, but it offers a much simpler analytical
representation.
In the limit of very large or very small temperatures, for instance,
the equivalent specific heat ratio reduces, respectively, to the
$4/3$ or $5/3$ limits.

The propagation of fast magneto-sonic shock waves has been
investigated by comparing the constant $\Gamma$ laws to the new
equation of state. Although for small shock velocities the shock dynamics
is well described by the cold gas limit, dynamical and thermodynamical quantities
(such as the compression ratio, internal energy, magnetization and so forth)
substantially change across the wave front at moderately
or highly relativistic speeds.
Eventually, for increasing shock velocities, flow quantities in the downstream
region smoothly vary from the cold ($\Gamma = 5/3$) to the hot ($\Gamma = 4/3$) regimes.

We numerically studied the effect of the EoS on shocks, blast waves,
the propagation of relativistic jets, and magnetized accretion flows
around Kerr black holes. Our results should serve as a useful guide
for future more specific studies of each topic.
For these numerical studies, we formulated the inversion from conservative
quantities to primitive quantities that allows a general EoS and avoids
catastrophic numerical cancellation in the non-relativistic and
ultrarelativistic limits. The analytical and numerical models confirm
the general result that large temperature gradients cannot be properly
described by a polytropic EoS with constant specific heat ratio.
Indeed, when compared to a more realistic EoS, for which the
polytropic index is a function of the temperature, considerable
dynamical differences arises.
This has been repeatedly shown in presence of strong discontinuities,
such shocks, across which the internal energy can change by several
order of magnitude.

We also showed that the turbulent behavior of magnetized accretion
flows around Kerr black holes depends on the EoS.  The $\Gamma=4/3$
EoS leads to more vigorous turbulence than the $\Gamma=5/3$ or TM
EoSs.  This affects the thickness of the corona that confines the
magnetized jet.  Any study of turbulence within the accretion disk,
the subsequent generation of heat in the coronae, and the
opening and acceleration of the jet (especially at large radii where
the cumulative differences due to the EoS in the disc are largest)
should use an accurate EoS.  The effect of the EoS on the jet opening angle
and Lorentz factor at large radii is a topic of future study.

The proposed equation state holds in the limit where effects due to
radiation pressure, electron degeneracies and neutrino physics
can be neglected.
It also omits potentially crucial physical aspects related to
kinetic processes (such as suprathermal particle distributions, cosmic rays),
plasma composition, turbulence effects at the sub-grid levels, etc.
These are very likely to alter the equation of state by effectively
changing the adiabatic index computed on merely thermodynamic
arguments.
Future efforts should properly address additional physical issues
and consider more general equations of state.

\section*{Acknowledgments}

We are grateful to our referee, P. Hughes, for his worthy
considerations and comments that led to the final form of this paper.
JCM was supported by a Harvard CfA Institute for Theory and
Computation fellowship.
AM would like to thank S. Massaglia and G. Bodo for useful
discussions on the jet propagation and morphology.

%
%

\appendix
\section{Primitive Variable Inversion Scheme}
\label{sec:inversion}
%
%
%
%
%

We outline a new primitive variable inversion scheme that is
used to convert the evolved conserved quantities into so-called
primitive quantities that are necessary to obtain the fluxes used for
the evolution. This scheme allows a general EoS by only
requiring specification of thermodynamical quantities and it also
avoids catastrophic cancellation in the non-relativistic and
ultrarelativistic limits.  Large Lorentz factors (up to $10^6$)
may not be uncommon in some
astrophysical contexts (e.g. Gamma-Ray-Burst) and ordinary inversion
methods can lead to severe numerical problems such as effectively
dividing by zero and subtractive cancellation, see, for instance,
\cite{BH06}.

First, we note that the general relativistic conservative quantities
can be written more
like special relativistic quantities by choosing a special frame in
which to measure all quantities.  A useful frame is the zero angular momentum
(ZAMO) observer in an axisymmetric space-time.  See~\citet{Noble06} for details.
From their expressions, it is useful to note that catastrophic
cancellations for non-relativistic velocities can be avoided by
replacing $\gamma-1$ in any expression with $(u_\alpha u^\alpha)/(\gamma+1)$,
where here $u_\alpha$ is the relative 4-velocity in the ZAMO frame.
From here on the expressions are in the ZAMO frame and appear similar
to the same expressions in special
relativity.

\subsection{Inversion Procedure}
%
%
%
%
%

Numerical integration of the conservation law (\ref{eq:CL}) proceeds
by evolving the conservative state vector $\vec{U} = \left(D,
\vec{m},\vec{B}, E\right)$ in time.  Computation of the fluxes,
however, requires velocity and pressure to be recovered from $\vec{U}$
by inverting Eqns.  (\ref{eq:cons_var_D})--(\ref{eq:cons_var_E}), a
rather time consuming and challenging task. For the constant-$\Gamma$
law, a recent work by~\citet{Noble06} examines several methods of
inversion.
In this section we discuss how to modify the equations of motion,
intermediate calculations, and the inversion from conservative to
primitive quantities so that the RMHD method 1) permits a general
EoS; and 2) avoids catastrophic cancellations in the
non-relativistic and ultrarelativistic limits.


Our starting relations are the total energy density (\ref{eq:cons_var_E}),
\begin{equation}\label{eq:energy}
E = W - p + \frac{1 + |\vec{v}|^2}{2} |\vec{B}|^2
            - \frac{S^2}{2 W^2}    \,,
\end{equation}
and the square modulus of Eq. (\ref{eq:cons_var_m}),
\begin{equation}\label{eq:mom2}
|\vec{m}|^2 =  \left(W + |\vec{B}|^2\right)^2|\vec{v}|^2  -
           \frac{S^2}{W^2} \left(2W + |\vec{B}|^2\right) \,,
\end{equation}
where $S\equiv \vec{m}\cdot\vec{B}$ and $W = Dh\gamma$.  Note that in order
for this expression to be accurate in the non-relativistic limit, one should
analytically cancel any appearance of $E$ in this expression.
Eq. (\ref{eq:mom2}) can be inverted to express the square of the
velocity in terms of the only unknown $W$:
\begin{equation}\label{eq:v2}
 |\vec{v}|^2 = \frac{S^2(2W + |\vec{B}|^2) + |\vec{m}|^2W^2}
                    {(W + |\vec{B}|^2)^2W^2} \,.
\end{equation}
After inserting (\ref{eq:v2}) into (\ref{eq:energy}) one has:
\begin{equation}\label{eq:energy_v2}
E = W - \pg + \frac{|\vec{B}|^2}{2} + \frac{|\vec{B}|^2 |\vec{m}|^2 - S^2}
                                            {2(|\vec{B}|^2 + W)^2} \,.
\end{equation}

In order to avoid numerical errors in the non-relativistic
limit one must modify the equations of motion and several intermediate
calculations. One solves the conservation equations with the mass
density subtracted from the energy by defining a new conserved
quantity ($E'=E-D$) and similarly for the energy flux. In
addition, operations based upon $\gamma$ can lead to catastrophic
cancellations since the residual $\gamma-1$ is often requested and is
dominant in the non-relativistic limit.  A more natural quantity to
consider is $|\vec{v}|^2$ or $\gamma^2 |\vec{v}|^2$.
Also, in the ultrarelativistic limit calculations based upon $\gamma(|\vec{v}|^2)$
have catastrophic cancellation errors when $|\vec{v}|\to 1$. This can be avoided by 1) using
instead $|\vec{u}|^2 \equiv \gamma^2 |\vec{v}|^2$ and 2) introducing the
quantities $E' = E-D$ and $W' = W - D$, with $W'$ properly rewritten as
\begin{equation}\label{eq:Wprime}
W'= \frac{D|\vec{u}|^2}{1 + \gamma} + \chi\gamma^2
\end{equation}
to avoid machine accuracy problems in the nonrelativistic limit, where
$\chi \equiv \rho\epsilon + \pg$. Thus our relevant equations become:
\begin{equation}\label{eq:main_Eprime}
E' = W' - \pg + \frac{|\vec{B}|^2}{2}
              + \frac{|\vec{B}|^2 |\vec{m}|^2 - S^2}{2(|\vec{B}|^2 + W'
              + D)^2} \,,
\end{equation}
\begin{equation}\label{eq:main_m2}
|\vec{m}|^2 = (W + |\vec{B}|^2)^2 \frac{|\vec{u}|^2}{1 + |\vec{u}|^2}
               - \frac{S^2}{W^2}\left(2W + |\vec{B}|^2\right) \,,
\end{equation}
where $W = W'+D$.

Equations (\ref{eq:main_Eprime}) and (\ref{eq:main_m2}) may be inverted
to find $W'$, $\pg$ and $|\vec{u}|^2$.
A one dimensional inversion scheme is derived by regarding
Eq. (\ref{eq:main_Eprime}) as a single nonlinear equation
in the only unknown $W'$ and using Eq. (\ref{eq:main_m2}) to express
$|\vec{u}|^2$ as a function of $W'$.
Using Newton's iterative scheme as our root finder, one needs
to compute the derivative
\begin{equation}\label{eq:dfdw}
\frac{dE}{dW'} = 1 - \frac{d\pg}{dW'}
                      - \frac{\left(|\vec{B}|^2 |\vec{m}|^2 - S^2\right)}{(|\vec{B}|^2 + W' + D)^3}\,.
\end{equation}
The explicit form of $d\pg/dW'$ depends on the particular EoS being
used.  While prior methods in principle allow for a general EoS, one
has to re-derive many quantities that involve kinematical
expressions. This can be avoided by splitting the kinematical and
thermodynamical quantities. This also allows one to write the
expressions so that there is no catastrophic cancellations in the
non-relativistic or ultrarelativistic limits.  Assuming that
$p=p(\chi,\rho)$, we achieve this by applying the chain rule to the
pressure derivative:
\begin{equation}\label{eq:press_splitting}
  \left(\frac{d\pg}{dW'}\right) =
  \pd{\pg}{\chi}\frac{d\chi}{dW'} + \pd{\pg}{\rho}\frac{d\rho}{dW'}\,.
\end{equation}
Partial derivatives involving purely thermodynamical quantities
must now be supplied by the EoS routines.
Derivatives with respect to $W'$, on the other hand, involve purely
kinematical terms and do not depend on the choice of the EoS.
Relevant expressions needed in our computations are given in the
Appendix.

Once $W'$ has been determined to some accuracy, the inversion process
is completed by computing the velocities from an inversion of equation
(\ref{eq:cons_var_m}) to obtain
\begin{equation}\label{velocityfromW}
  v_k = \frac{1}{W + |\vec{B}|^2}\left(m_k + \frac{S}{W} B_k\right) \,,
\end{equation}
One then computes $\chi$ from an inversion
of equation (\ref{eq:Wprime}) to obtain
\begin{equation}\label{eq:chi}
\chi = \frac{W'}{\gamma^2} - \frac{D|\vec{u}|^2}{(1 + \gamma)\gamma^2} ,
\end{equation}
from which $\pg$ or $\rho\epsilon$ can be obtained for any given EoS.
The rest mass density is obtained from
\begin{equation}
 \rho = \frac{D}{\gamma} \,,
\end{equation}
and the magnetic field is trivially inverted.

In summary, we have formulated an inversion scheme that 1) allows a
general EoS without re-deriving kinematical expressions; and 2) avoids
catastrophic cancellation in the non-relativistic and
ultrarelativistic limits.  This inversion involves solving a single
non-linear equation using, e.g., a one-dimensional Newton's method.  A
similar two-dimensional method can be easily written with the same
properties, and such a method may be more robust in some cases since
the one-dimensional version described here involves more complicated
non-linear expressions.

One can show analytically that the inversion is accurate
in the ultrarelativistic limit as long as
$\gamma\lesssim \epsilon^{-1/2}_{\rm machine}$
for $\gamma$ and $p/(\rho\gamma^2)\gtrsim \epsilon_{\rm machine}$ for pressure,
where $\epsilon_{\rm machine}\approx 2.2\times 10^{-16}$ for double precision.
The method used by~\citet{Noble06} requires $\gamma\lesssim \epsilon^{-1/2}_{\rm machine}/10$
due to the repeated use of the expression $\gamma=1/\sqrt{1-v^2}$ in the inversion.
Note that we use $\gamma=\sqrt{1+|u|^2}$ that has no catastrophic cancellation.
The fundamental limit on accuracy is due to evolving energy and momentum separately
such that the expression $E-|m|$ appears in the inversion.  Only a method that evolves
this quantity directly (e.g. for one-dimensional problems one can evolve the
energy with momentum subtracted) can reach higher Lorentz factors.
An example test problem is the ultrarelativistic Noh test in~\citet{alo99} with
$p=7.633\times 10^{-6}$, $\Gamma=4/3$, $1-v=10^{-11}$ (i.e. $\gamma=223607$)
This test has $p/(\rho\gamma^2)\approx 1.6\times 10^{-16}$, which is just below
double precision and so the pressure is barely resolved in the pre-shock region.  The
post-shock region is insensitive to the pre-shock pressure and
so is evolved accurately up to $\gamma\approx 6\times 10^7$.  These facts are
have been also confirmed numerically using this inversion within HARM.
Using the same error measures as in~\citet{alo99} we can evolve their test problem
with an even higher Lorentz factor of $\gamma=10^7$ and obtain similar errors
of $\lesssim 0.1\%$.

\subsection{Kinematical and Thermodynamical Expressions}
%
%

The kinematical terms required in equation (\ref{eq:press_splitting})
may be easily found from the definition of $W'$,
\begin{equation}
  W' \equiv Dh\gamma - D = D(\gamma-1) + \chi\gamma^2 \,,
\end{equation}
by straightforward differentiation. This yields
\begin{equation}\label{eq:dchidw}
   \frac{d\chi}{dW'} = \frac{1}{\gamma^2}
 - \frac{\gamma}{2}(D + 2\gamma\chi)\frac{d|\vec{v}|^2}{dW'}  \,,
\end{equation}
and
\begin{equation}\label{eq:drhodw}
   \frac{d\rho}{dW'} = D\frac{d(1/\gamma)}{dW'} = -\frac{D\gamma}{2}
\frac{d|\vec{v}|^2}{dW'}  \,,
\end{equation}
where
\begin{equation}
 \frac{d|\vec{v}|^2}{dW} =
 -\frac{2}{W^3}\frac{S^2\left[3W(W + |\vec{B}|^2) + |\vec{B}|^4\right]
                     + |\vec{m}|^2W^3}
  {\left(W + |\vec{B}|^2\right)^3}   \,,
\end{equation}
is computed by differentiating (\ref{eq:v2}) with respect to $W$
(note that $d/dW' \equiv d/dW$).
Equation (\ref{eq:dchidw}) does not depend on the
knowledge of the EoS.

Thermodynamical quantities such as $\partial p/\partial\chi$,
on the other hand, do require the explicit form of the EoS.
For the ideal gas EoS one simply has
\begin{equation}\label{eq:pchi_id}
  \pg(\chi,\rho) = \frac{\Gamma-1}{\Gamma} \chi \,,
\end{equation}
where $\chi = \rho\epsilon + p$.
By taking the partial derivatives of (\ref{eq:pchi_id})
with respect to $\chi$ (keeping $\rho$ constant) and
$\rho$ (keeping $\chi$ constant) one has
\begin{equation}
  \pd{\pg}{\chi} = \frac{\Gamma-1}{\Gamma} \;,\quad
  \pd{\pg}{\rho} = 0  \,.
\end{equation}

For the TM EoS, one can more conveniently rewrite (\ref{eq:TM_int_en})
as
\begin{equation}
  3p(\rho + \chi - p) = (\chi-p)(\chi + 2\rho - p)\,,
\end{equation}
which, upon differentiation with respect to $\chi$ (keeping $\rho$ constant)
yields
\begin{equation}
\pd{p}{\chi} = \frac{2\chi + 2\rho - 5p}{5\rho + 5\chi - 8p}\,.
\end{equation}
Similarly, by taking the derivative with respect to $\rho$ at
constant $\chi$ gives
\begin{equation}
 \pd{p}{\rho} = \frac{2\chi - 5p}{5\rho + 5\chi - 8p} \,.
\end{equation}

In order to use the above expressions and avoid catastrophic
cancellation in the non-relativistic limit, one must solve for the gas
pressure as functions of only $\rhorest$ and $\chi$ and then write the
pressure that explicitly avoids catastrophic cancellation as
$\{\chi,p\}\to 0$.  One obtains:
\begin{equation}
  \pg(\chi,\rhorest)  = \frac{2\chi(\chi + 2\rho)}{5(\chi + \rho) + \sqrt{9\chi^2+18\rho\chi+25\rho^2}} .
\end{equation}
Also, for setting the initial conditions it is useful to be able to
convert from a given pressure to the internal energy by using
\begin{equation}
  \rhorest\epsilon(\rhorest,\pg) =
  \frac{3}{2}\left(\pg + \frac{3\pg^2}{2\rhorest+\sqrt{9\pg^2+4\rhorest^2}}\right) ,
\end{equation}
which also avoids catastrophic cancellation in the non-relativistic limit.

\subsection{Newton-Raphson Scheme}
%
%
%
%

Equation (\ref{eq:main_Eprime}) may be solved using a Newton-Raphson iterative
scheme, where the $(k+1)$-th approximation to the $W'$ is computed as
\begin{equation}\label{eq:newton}
  W'^{(k+1)} = W'^{(k)} - \left.\frac{f(W')}{df(W')/dW'}\right|_{W'=W'^{(k)}} \,,
\end{equation}
where
\begin{equation}
 f(W') = W' - E' - \pg + \frac{|\vec{B}|^2}{2} + \frac{|\vec{B}|^2 |\vec{m}|^2 - S^2}{2(|\vec{B}|^2 + W' + D)^2} \,,
\end{equation}
and $df(W')/dW' \equiv dE'/dW'$ is given by Eq. (\ref{eq:dfdw}).
The iteration process terminates when the residual
$\left|W'^{(k+1)}/W'{(k)} - 1\right|$
falls below some specified tolerance.

We remind the reader that, in order to start the iteration process
given by (\ref{eq:newton}), a suitable initial guess must be provided.
We address this problem by initializing, at the beginning of the
cycle, $W'^{(0)} = \tilde{W}_+ - D$, where $\tilde{W}_+$ is the positive
root of
\begin{equation}
  {\cal P}(W,1) = 0 \,,
\end{equation}
and ${\cal P}(W,|\vec{v}|)$ is the quadratic function
\begin{equation}
  {\cal P}(W,|\vec{v}|) = |\vec{m}|^2 - |\vec{v}|^2 W^2
           + (2W + |\vec{B}|^2)(2W + |\vec{B}|^2 - 2E)\,.
\end{equation}
This choice guarantees positivity of pressure, as it can be proven
using the relation
\begin{equation}\label{eq:pressW2}
  p = \frac{{\cal P}(W,|\vec{v}|)}{2(2W + |\vec{B}|^2)}\,,
\end{equation}
which follows upon eliminating the $(S/W)^2$ term in
Eq. (\ref{eq:mom2}) with the aid of Eq. (\ref{eq:energy}).  Seeing that
${\cal P}(W,|\vec{v}|)$ is a convex quadratic function, the condition $p>0$ is
equivalent to the requirement that the solution $W$ must lie outside
the interval $[W_-,W_+]$, where ${\cal P}(W_\pm,|\vec{v}|)=0$.  However, since
${\cal P}(W,|\vec{v}|) \ge {\cal P}(W,1)$, it must follow that $\tilde{W}_+ \ge W_+$ and
thus $\tilde{W}_+$ lies outside the specified interval. We tacitly
assume that the roots are always real, a condition that is always met
in practice.

\label{lastpage}
\end{document}